\documentclass[cite,aps,pra,amsmath, amsfonts, amssymb, superscriptaddress, twocolumn, showpacs]{revtex4}

\usepackage{graphicx}
\usepackage{setspace}

\renewcommand{\vec}[1]{\mathbf{#1}} 

\newcommand{\citeasnoun}[1]{Ref.~\onlinecite{#1}}

\newcommand{\chiany}[1]{\chi^{(#1)}}
\newcommand{\chitwo}{\chiany{2}}
\newcommand{\chithree}{\chiany{3}}

\renewcommand{\eqref}[1]{Eq.~(\ref{#1})}

\newcommand{\eqrefrange}[2]{Eqs.~(\ref{#1}--\ref{#2})}

\newcommand{\secref}[1]{Sec.~\ref{sec:#1}}

\newcommand{\figref}[1]{Fig.~\ref{fig:#1}}
\newcommand{\Figref}[1]{Figure~\ref{fig:#1}}
\newcommand{\figreftwo}[2]{Figs.~\ref{fig:#1} and~\ref{fig:#2}}

\newcommand{\figrefrange}[2]{Figs.~\ref{fig:#1}--\ref{fig:#2}}

\begin{document}
\def\linefigwidth{0.5\textwidth}
\def\smalllinefigwidth{0.35\textwidth}
\def\smallerlinefigwidth{0.25\textwidth}
\def\largelinefigwidth{0.5\textwidth}

\title{Nonlinear harmonic generation and devices in doubly-resonant Kerr cavities}



\author{Hila Hashemi}
\affiliation{Department of Mathematics, Massachusetts Institute of Technology, Cambridge, MA 02139}
\author{Alejandro W. Rodriguez}
\affiliation{Department of Physics, Massachusetts Institute of Technology, Cambridge, MA 02139}
\author{J.~D.~Joannopoulos}
\affiliation{Department of Physics, Massachusetts Institute of Technology, Cambridge, MA 02139}
\author{Marin~Solja{\v{c}}i{\'{c}}}
\affiliation{Department of Physics, Massachusetts Institute of Technology, Cambridge, MA 02139}
\author{Steven~G.~Johnson}
\affiliation{Department of Mathematics, Massachusetts Institute of Technology, Cambridge, MA 02139}

\begin{abstract}
  We describea theoretical analysis of the nonlinear dynamics of third-harmonic   generation ($\omega\to3\omega$) via Kerr ($\chithree$) nonlinearities in a
  resonant cavity with resonances at both $\omega$ and $3\omega$.
  Such a doubly resonant cavity greatly reduces the required power for
  efficient harmonic generation, by a factor of $\sim V/Q^2$ where $V$
  is the modal volume and $Q$ is the lifetime, and can even exhibit
  100\% harmonic conversion efficiency at a critical input power.
  However, we show that it also exhibits a rich variety of nonlinear
  dynamics, such as multistable solutions and long-period limit
  cycles.We describe how to compensate for self/cross-phase modulation (which
  otherwise shifts the cavity frequencies out of resonance), and how
  to excite the different stable solutions (and especially the
  high-efficiency solutions) by specially modulated input pulses.
\end{abstract}

\maketitle 


\section{Introduction}

In this paper, we describe how 100\% third-harmonic conversion can
occur in doubly-resonant optical cavities with Kerr nonlinearities,
even when dynamical stability and self-phase modulation (which can
drive the cavities out of resonance) are included (extending our
earlier work~\cite{Rodriguez07:OE}), and describe the initial
conditions required to excite these efficient solutions.  In
particular, we show that such doubly-resonant nonlinear optical
systems can display a rich variety of dynamical behaviors, including
multistability (different steady states excited by varying initial
conditions, a richer version of the bistable phenomenon observed in
single-mode cavities~\cite{Soljacic02:bistable}), long-period limit
cycles (similar to the ``self-pulsing'' observed for second-harmonic
generation~\cite{Drummond80}), and transitions in the stability and
multiplicity of solutions as the parameters are varied.  One
reason for studying such doubly resonant cavities was the fact that
they lower the power requirements for nonlinear
devices~\cite{Rodriguez07:OE}, and in particular for third harmonic
conversion, compared to singly-resonant cavities or nonresonant
structures~\cite{Armstrong62,Ashkin66,Smith70,Ferguson77,Brieger81,Berquist82,Kozlovsky88,Dixon89,Collet90,Persaud90,Moore95,Schneider96,Mu01,Hald01,McConnell01,Dolgova02,Liu05,Scaccabarozzi06}.
An appreciation and understanding of these behaviors is important to
design efficient harmonic converters (the main focus of this paper),
but it also opens the possibility of new types of devices enabled by
other aspects of the nonlinear dynamics.

\begin{figure}[t]
\centering
\includegraphics[width=0.50\textwidth]{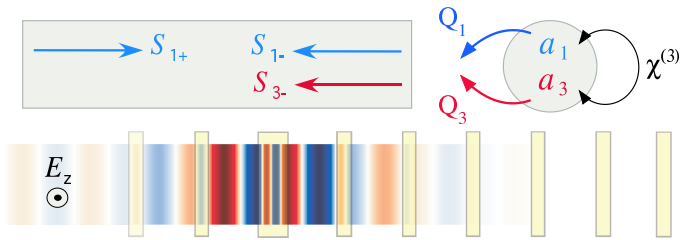}
\caption{\emph{Top:} Schematic of general scheme for third-harmonic
generation, and dynamical variables for coupled-mode equations: a
single input/output channel (with incoming/outgoing field amplitudes
$s_\pm$) is coupled to a resonant cavity with two modes at frequencies
$\omega_1$ and $3\omega_1$ (and corresponding amplitudes $a_1$ and
$a_3$).  The two resonant modes are nonlinearly coupled by a Kerr
($\chithree$) nonlinearity.  \emph{Bottom:} An example realization~\cite{Rodriguez07:OE}, in
one dimension, using a semi-infinite quarter-wave stack of dielectric
layers with a doubled-layer defect (resonant cavity) that is coupled
to incident plane waves; the electric field of a steady-state $3\omega_1$ solution
is shown as blue/white/red for negative/zero/positive.}
\label{fig:schematic}
\end{figure}

In a Kerr ($\chithree$) medium, there is a change in the refractive
index proportional to the square of the electric field; for an
oscillating field at a frequency $\omega$, this results in a shift in
the index at the same frequency (self-phase modulation, SPM),
generation of power at the third-harmonic frequency $3\omega$, and
also other effects when multiple frequencies are present [cross-phase
modulation (XPM) and four-wave mixing (FWM)]~\cite{Boyd92}.  When the
field is confined in a cavity, restricting to a small modal volume $V$
for a long time given by the quality factor $Q$ (a lifetime in units
of the optical period)~\cite{JoannopoulosJo08-book}, such nonlinear
effects are enhanced by both the increased field strength for the same
input power and by the frequency sensitivity inherent in resonant
effects (since the fractional bandwidth is $1/Q$).  This enhancement
is exploited, for example, in nonlinear harmonic and sum-frequency
generation, most commonly for $\chitwo$ effects where the change in
index is proportional to the electric field (which requires a
non-centrosymmetric material)~\cite{Boyd92}.  One can further enhance
harmonic generation by using a cavity with \emph{two} resonant modes,
one at the source frequency and one at the harmonic
frequency~\cite{Paschotta94,Berger96,Zolotoverkh00,Maes05,Liscidini06,Dumeige06,Drummond80,
Wu87, Ou93}.  In this case, one must also take into account a
nonlinear downconversion process that competes with harmonic
generation~\cite{Drummond80, Wu87, Ou93}, but it turns out to be
theoretically possible to obtain 100\% harmonic conversion for either
$\chitwo$ ($\omega\to2\omega$) or $\chithree$ ($\omega\to3\omega$)
nonlinearities at a specific ``critical'' input power
$P_\mathrm{crit}$ (both in a one-dimensional model of propagating
waves for $\chitwo$ nonlinearities~\cite{Schiller93:PhD} and also in a
more general coupled-mode model for either $\chitwo$ or $\chithree$
nonlinearities~\cite{Rodriguez07:OE}).  In particular, we studied the
harmonic-generation and downconversion processes in a broad class of
model systems depicted in \figref{schematic}: a single input channel
(e.g. a waveguide) is coupled to a nonlinear cavity with two resonant
frequencies, where both reflected and harmonic fields are emitted back
into the input channel.  In this case, we predicted 100\% harmonic
generation at a critical power $P_\mathrm{crit}$ proportional to $V /
Q^3$ for $\chitwo$ and $V/Q^2$ for $\chithree$~\cite{Rodriguez07:OE}.
However, we only looked at the steady-state solution of the system and
not its dynamics or stability.  Moreover, in the $\chithree$ case
there can also be an SPM/XPM effect that shifts the cavity frequencies
out of resonance and spoils the harmonic-generation effect.  In this
paper, we consider both of these effects, describe how to compensate
for SPM/XPM, and demonstrate the different regimes of stability in
such $\chithree$ doubly resonant systems.  We show that the parameters
and the initial conditions must be chosen within certain regimes to
obtain a stable steady state with high conversion efficiency.

In other regimes, we demonstrate radically different behaviors: not
only low-efficiency steady states, but also limit-cycle solutions
where the efficiency oscillates slowly with a repetition period of
many thousands of optical cycles.  With infrared light, these limit
cycles form a kind of optical oscillator/clock with a period in the
hundreds of GHz or THz (and possibly lower, depending on the cavity
parameters).  Previously, limit-cycle/self-pulsing behaviors have been
observed in a number of other nonlinear optical systems, such as:
doubly-resonant $\chitwo$ cavities coupled by second-harmonic
generation~\cite{Drummond80}; bistable multimode Kerr cavities with
time-delayed nonlinearities~\cite{Abraham82}; nonresonant distributed
feedback in Bragg gratings~\cite{Parini07}; and a number of nonlinear
lasing devices~\cite{Siegman86}.  However, the system considered in
this work seems unusually simple, especially among $\chithree$
systems, in that it only requires two modes and an instantaneous Kerr
nonlinearity, with a constant-frequency input source, to attain
self-pulsing, and partly as a consequence of this simplicity the
precise self-pulsing solution is quite insensitive to the initial
conditions.  In other nonlinear optical systems where self-pulsing was
observed, other authors have also observed chaotic solutions in
certain regimes.  Here, we did not observe chaos for any of the
parameter regimes considered, where the input was a constant-frequency
source, but it is possible that chaotic solutions may be excited by an
appropriate pulsed input as in the $\chitwo$ case~\cite{Drummond80}.

Another interesting phenomenon that can occur in nonlinear systems is
multistability, where there are multiple possible steady-state
solutions that one can switch among by varying the initial conditions.
In Kerr ($\chithree$) media, an important example of this phenomenon
is bistable transmission through nonlinear cavities: for transmission
through a \emph{single}-mode cavity, output can switch discontinuously
between a high-transmission and a low-transmission state in a
hysteresis effect that results from SPM~\cite{Soljacic02:bistable}.
For example, if one turns on the power gradually from zero the system
stays in the low-transmission state, but if the power is increased
further and then decreased to the original level, the system can be
switched to the high-transmission state.  This effect, which has been
observed experimentally~\cite{Notomi05}, can be used for all-optical
logic, switching, rectification, and many other
functions~\cite{Soljacic02:bistable}.  In a cavity with multiple
closely-spaced resonances, where the nonlinearity is strong enough to
shift one cavity mode's frequency to another's, the same SPM
phenomenon can lead to more than two stable solutions~\cite{Felber76}.
Here, we demonstrate a much richer variety of multistable phenomena in
the doubly-resonant case for widely-separated cavity frequencies
coupled by harmonic generation in addition to SPM---not only can there
be more than two stable states, but the transitions between them can
exhibit complicated oscillatory behaviors as the initial conditions
are varied, and there are also Hopf bifurcations into self-pulsing
solutions.

The remaining part of the paper is structured as follows.  In \secref{theory}, we review the theoretical description of harmonic generation in a doubly
resonant cavity coupled to input/output waveguides, based on temporal
coupled-mode theory, and demonstrate the possibility of 100\% harmonic
conversion.  We also discuss how to compensate for frequency shifting
due to SPM and XPM by pre-shifting the cavity frequencies.  In
\secref{stability}, we analyze the stability of this 100\%-efficiency
solution, and demonstrate the different regimes of stable operation
that are achieved in practice starting from that theoretical initial
condition.  We also present bifurcation diagrams that show how the
stable and unstable solutions evolve as the parameters vary.  Finally,
in \secref{exciting}, we consider how to excite these high-efficiency
solutions in practice, by examining the effect of varying initial
conditions and uncertainties in the cavity parameters.  In particular,
we demonstrate the multistable phenomena exhibited as the initial
conditions are varied.  We close with some concluding remarks,
discussing the many potential directions for future work that are
revealed by the phenomena described here.

\section{100\% harmonic conversion in doubly-resonant cavities}
\label{sec:theory}

In this section, we describe the basic theory of frequency conversion in
doubly-resonant cavities with $\chithree$ nonlinearities, including
the undesirable self- and cross-phase modulation effects, and explain
the existence of a solution with 100\% harmonic conversion (without
considering stability).

Consider a waveguide coupled to a doubly resonant cavity with two
resonant frequencies $\omega_1^\mathrm{cav} = \omega_1$ and
$\omega_3^\mathrm{cav} = \omega_3 = 3\omega_1$ (below, we will shift
$\omega_k^\mathrm{cav}$ to differ slightly from $\omega_k$), and
corresponding lifetimes $\tau_1$ and $\tau_3$ describing their
radiation rates into the waveguide (or quality factors $Q_k = \omega_k
\tau_k / 2$).  In addition, these modes are coupled to one another via
the Kerr nonlinearity.  Because all of these couplings are weak, any
such system (regardless of the specific geometry), can be accurately
described by temporal coupled-mode theory, in which the system is
modeled as a set of coupled ordinary differential equations
representing the amplitudes of the different modes, with coupling
constants and frequencies determined by the specific
geometry~\cite{Suh04,Rodriguez07:OE}.  In particular, the coupled-mode
equations for this particular class of geometries were derived in
\citeasnoun{Rodriguez07:OE} along with explicit equations for the
coupling coefficients in a particular geometry.  The degrees of
freedom are the field amplitude $a_k$ of the $k$th cavity mode
(normalized so that $|a_k|^2$ is the corresponding energy) and the
field amplitude $s_{k\pm}$ of the incoming ($+$) and outgoing ($-$)
waveguide modes at $\omega_k$ (normalized so that $|s_{k\pm}|^2$ is
the corresponding power), as depicted schematically in
\figref{schematic}.  These field amplitudes are coupled by the
following equations (assuming that there is only input at $\omega_1$,
i.e. $s_{3+}=0$):
\begin{eqnarray}
\begin{split}
\dot{a_1} &=&\left [i\omega_1^\mathrm{cav} \left(1-\alpha_{11} |a_1|^{2}-\alpha_{13}|a_3|^{2}\right)-\frac{1}{\tau_{1}}\right] a_{1}\\
& &{} -i\omega_1 \beta_1(a_1^*)^2 a_3+\sqrt{\frac{2}{\tau_{s,1}}} s_{1+}
\label{gov1}
\end{split}
\\ \begin{split} \dot{a_3} &=& \left [i\omega_3^\mathrm{cav}(1-\alpha_{31}|a_1|^2-\alpha_{33} |a_3|^2)-\frac{1}{\tau_{3}}\right ]a_3\\
& &{ } -i\omega_3\beta_3a_1^{3}
\label{gov2}
\end{split}
\end{eqnarray}
As explained in \citeasnoun{Rodriguez07:OE}, the $\alpha$ and $\beta$
coefficients are geometry/material-dependent constants that express
the strength of various nonlinear effects for the given modes.  The
$\alpha_{ij}$ terms describe self- and cross-phase modulation effects:
they clearly give rise to effective frequency shifts in the two
modes. The $\beta_i$ term characterize the energy transfer between the
modes: the $\beta_3$ term describes frequency up-conversion and the
$\beta_1$ term describes down-conversion. As shown in
\citeasnoun{Rodriguez07:OE}, they are related to one another via
conservation of energy $\omega_{1}\beta_{1}=\omega_{3}\beta_{3}^{*}$,
and all of the nonlinear coefficients scale inversely with the modal
volume $V$.

There are three different $\alpha_{ij}$ parameters (two SPM
coefficients $\alpha_{11}$ and $\alpha_{33}$ and one XPM coefficient
$\alpha_{13} = \alpha_{31}$).  All three values are different, in
general, but are determined by similar integrals of the field
patterns, produce similar frequency-shifting phenomena, and all scale
as $1/V$.  Therefore, in order to limit the parameter space analyzed
in this paper, we consider the simplified case where all three
frequency-shifting terms have the same strength $\alpha_{ij} =
\alpha$.

One can also include various losses, e.g. linear losses correspond to
a complex $\omega_1$ and/or $\omega_3$, and nonlinear two-photon
absorption corresponds to a complex $\alpha$.  As discussed in the
concluding remarks, however, we have found that such considerations do
not qualitatively change the results (only reducing the efficiency
somewhat, as long as the losses are not too big compared to the
radiative lifetimes $\tau$), and so in this manuscript we restrict
ourselves to the idealized lossless case.

\begin{figure}[t]
 \centering \includegraphics[width=0.49\textwidth]{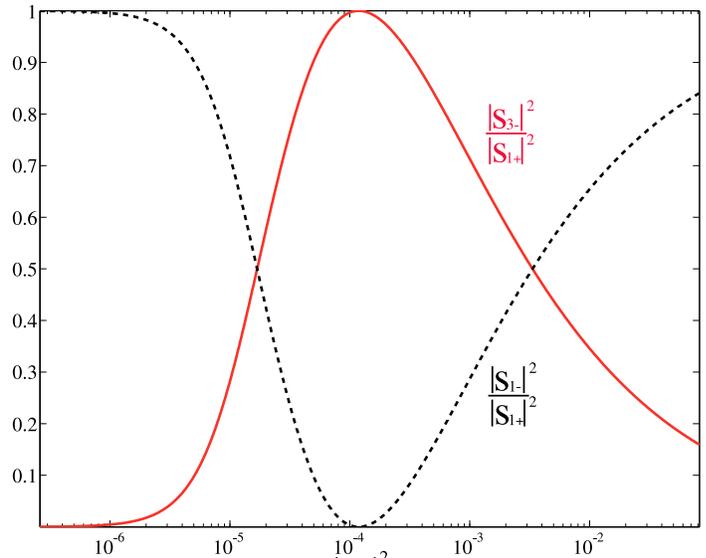}
 \caption{Steady-state efficiency of third-harmonic generation (solid
 red line) from \citeasnoun{Rodriguez07:OE}, for $\alpha=0$ (no
 self-phase modulation), as a function of input power $|s_{1+}|^2$
 scaled by the Kerr coefficient $n_2 = 3\chithree/4c\varepsilon$.  The
 reflected power at the incident frequency $\omega_1$ is shown as a
 dashed black line.  There is a critical power where the efficiency of
 harmonic generation is 100. The parameters used in this plot are $Q_{1}=1000$, $Q_{3}=3000$, $\beta_{1}=(4.55985-0.7244i)\times 10^{-5}$ in dimensionless units of $\chithree/V\epsilon$.}
 \label{fig:critpower}
\end{figure}

\Figref{critpower} shows the steady-state conversion efficiency
($|s_{3-}|^2/|s_{1+}|^2$) versus input power of light that is incident
on the cavity at $\omega^{cav}_1$, for the same parameter regime in
\citeasnoun{Rodriguez07:OE} (i.e. assuming negligible self- and
cross-phase modulation so that $\alpha = 0$), and not considering the
stability of the steady state.  As shown by the solid red curve, as
one increases the input power, the efficiency increases, peaking at
100\% conversion for a critical input power $P_\mathrm{crit} =
|s_{1+}^\mathrm{crit}|$, where
\begin{equation}
|s_{1+}^\mathrm{crit}| = \left( \frac{4}{|\omega_1 \beta_1|^2
 \tau_1^3 \tau_3} \right)^{1/4}.
\label{s1crit}
\end{equation}
The efficiency decreases if the power is either too low (in the linear regime) or
too high (dominated by down-conversion).  $P_\mathrm{crit}$ scales as
$V/Q^2$, so one can in principle obtain very low-power efficient
harmonic conversion by increasing $Q$ and/or decreasing
$V$~\cite{Rodriguez07:OE}.  Including absorption or other losses
decreases the peak efficiency, but does not otherwise qualitatively
change this solution~\cite{Rodriguez07:OE}.

\begin{figure}[t]
 \centering
 \includegraphics[width=0.5\textwidth]{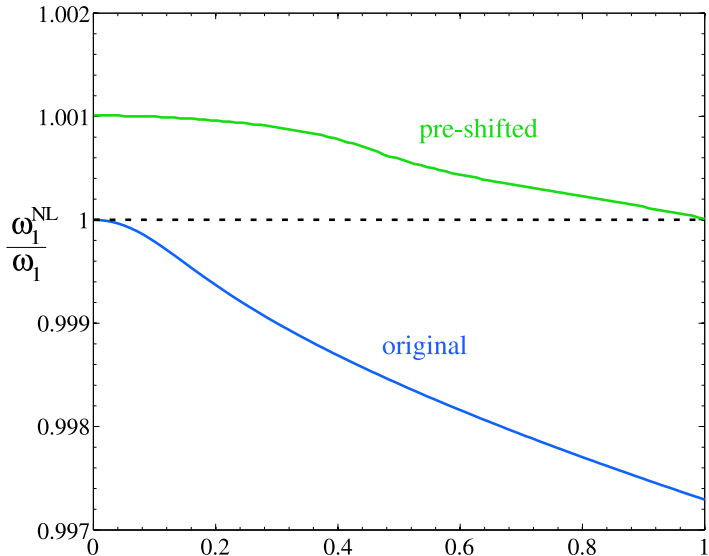}
 \caption{Shift in the resonant frequency $\omega_1^\mathrm{NL}$ as a
 function of input power, due to self/cross-phase modulation. (There
 is an identical shift in $\omega_3^\mathrm{NL}$.)  If the cavity is
 designed so that the linear ($P_\mathrm{in}\to0$) frequencies are
 harmonics, the nonlinearity pushes the system out of resonance (lower
 blue line) as the power increases to the critical power for 100\%
 efficiency. This is corrected by pre-shifting the cavity frequencies
 (upper green line) so that the nonlinear frequency shift pushes the
 modes into resonance at $P_\mathrm{crit}$.}
 \label{fig:detuning}
\end{figure}

There are two effects that we did not previously analyze in detail, however,
which can degrade this promising solution: nonlinear frequency shifts
and instability. Here, we first consider frequency shifts, which arise
whenever $\alpha\neq0$, and consider stability in the next
section.  The problem with the $\alpha$ terms is that efficient
harmonic conversion depends on the cavities being tuned to harmonic
frequencies $\omega_3 = 3\omega_1$; a nonlinear shift in the cavity
frequencies due to self/cross-phase modulation will spoil this
resonance.  In principle, there is a straightforward solution to this
problem, as depicted in \figref{detuning}.  Originally (for $\alpha=0$), the
cavity was designed to have the frequency $\omega_1$ in the linear
regime, but with $\alpha\neq0$ the effective cavity frequency
$\omega_1^\mathrm{NL}$ (including self/cross-phase modulation terms)
is shifted away from the design frequency as shown by the blue line.
Instead, we can simply design the linear cavity to have a frequency
$\omega_1^\mathrm{cav}$ slightly different from the operating
frequency $\omega_1$, so that self/cross-phase modulation shifts
$\omega_1^\mathrm{NL}$ exactly to $\omega_1$ at the critical input
power, as depicted by the green line in \figref{detuning}.  Exactly the same
strategy is used for $\omega_3^\mathrm{NL}$, by pre-shifting
$\omega_3^\mathrm{cav}$.

More precisely, to compute the required amount of pre-shifting, we
examine the coupled-mode equations \eqrefrange{gov1}{gov2}.  First, we solve for the
critical power $P_\mathrm{crit}$ assuming $\alpha=0$, as in
\citeasnoun{Rodriguez07:OE}, and obtain the corresponding critical cavity fields
$a_{k}^\mathrm{crit}$:
\begin{align}
\left|a_1^\mathrm{crit}\right|^2= \left(\frac{1}{\omega_1^2\left|\beta_1\right|^2\tau_3\tau_{1,s}}\right)^{1/2} \\
\left|a_3^\mathrm{crit}\right|^2= \left(\frac{\omega_3\beta_3\tau_3}{(\omega_1\beta_1\tau_{1,s})^3}\right)^{1/2}.
\end{align}
Then, we substitute these critical fields into the coupled-mode
equations for $\alpha \neq 0$, and solve for the new cavity
frequencies $\omega_k^\mathrm{cav}$ so as to cancel the $\alpha$ terms
and make the $a_{k}^\mathrm{crit}$ solutions still valid.  This yields
the following transformation of the cavity frequencies:
\begin{align}
\omega_{1}^\mathrm{cav} &=  \frac{\omega_{1}}{1-\alpha_{11}|a_{1}^\mathrm{crit}|^{2}-\alpha_{13}|a_{3}^\mathrm{crit}|^{2}} \\
\omega_{3}^\mathrm{cav} &= \frac{\omega_{3}}{1-\alpha_{13}|a_{1}^\mathrm{crit}|^{2}-\alpha_{33}|a_{3}^\mathrm{crit}|^{2}}.
\end{align}
By inspection, when substituted into \eqrefrange{gov1}{gov2} at the critical power,
these yield the same steady-state solution as for $\alpha=0$.  (There
are two other appearances of $\omega_1$ and $\omega_3$ in the
coupled-mode equations, in the $\beta_k$ terms, but we need not change
these frequencies because that is a higher-order effect, and the
derivation of the coupled-mode equations considered only first-order
terms in $\chithree$.)

The nonlinear dynamics turn out to depend only on four dimensionless
parameters: $\tau_3/\tau_1 = Q_3/3Q_1$, $\alpha_{11}/\beta_1$,
$\alpha_{33}/\beta_1$, and $\alpha_{13}/\beta_1 =
\alpha_{31}/\beta_1$.  The overall scale of $Q$, $\alpha$, etcetera,
merely determines the absolute scale for the power requirements: it is
clear from the equations that multiplying all $\alpha$ and $\beta$
coefficients by an overall constant $K$ can be compensated by dividing
all $a$ and $s$ amplitudes by $\sqrt{K}$ [which happens automatically
for $s$ at the critical power by \eqref{s1crit}]; the case of scaling
$\tau_{1,3}$ by an overall constant is more subtle and is considered
below. As mentioned above, for simplicity we take
$\alpha_{11}=\alpha_{33}=\alpha_{13}=\alpha_{31}=\alpha$.  Therefore,
in the subsequent sections we will analyze the asymptotic efficiency
as a function of $\tau_3/\tau_1$ and $\alpha/\beta_1$.

So far, we have found a steady-state solution to the coupled-mode
equations, including self/cross-phase modulation, that achieves 100\%
third-harmonic conversion.  In the following sections, we consider
under what conditions this solution is stable, what other stable
solutions exist, and for what initial conditions the high-efficiency
solution is excited.

To understand the dynamics and stability of the nonlinear coupled-mode
equations, we apply the standard technique of identifying fixed points
of the equations and analyzing the stability of the linearized
equations around each fixed point~\cite{Tabor89}.

By a ``fixed point,'' we mean a steady-state solution corresponding to
an input frequency $\omega_1$ ($s_{1+} \sim e^{-i\omega_1 t}$) and
hence $a_1(t) = A_1 e^{-i\omega_1 t}$ and $a_3(t) = A_3 e^{-i3\omega_1
t}$ for some unknown constants $A_1$ and $A_3$. [An input frequency
$\omega_1$ can also generate higher harmonics, such as $9\omega_1$ or
$5\omega_1$, but these are negligible: both because they are
higher-order effects ($\sim [\chithree]^2$, and all such terms were
dropped in deriving the coupled-mode equations), and because we assume there is
no resonant mode present at those frequencies.]  By substituting this
steady-state form into \eqrefrange{gov1}{gov2}, one obtains two
coupled polynomial equations whose roots are the fixed points.  We
already know one of the fixed points from the previous section, the
100\% efficiency solution, but to fully characterize the system one
would like to know all of the fixed points (both stable and unstable).
We solved these polynomial equations using Mathematica, which is able
to compute all of the roots, but some transformations were required to
put the equations into a solvable form.  In particular, we eliminated
the complex conjugations by writing $A_k = r_k e^{i\phi_k}$ and
assuming (without loss of generality) that $s_{1+}$ is
real. Multiplying \eqref{gov1} by $e^{-i\phi_1}$ and \eqref{gov2} by
$e^{-i3\phi_1}$ allows us to simply solve the system for
$e^{-i\phi_1}$ and $e^{i(\phi_3-3\phi_1)}$. Requiring the magnitude of
these two quantities to be unity yields two polynomials in $x=r_1^2$
and $y=r_3^2$, which Mathematica can handle. The resulting polynomial
is of an artificially high degree, resulting in spurious roots, but
the physical solutions are easily identified by the fact that $x$ and
$y$ must be real and non-negative. (We should also note that this
root-finding process is highly sensitive to roundoff
error~\cite{Press92}, independent of the physical stability of the
solutions, but we dealt with that problem by employing 50 decimal
places of precision.)

As mentioned above, the dynamics are independent of the overall scale
of $\tau_{1,3}$, and depend only on the ratio $\tau_3/\tau_1$.  This
can be seen from the equations for $A_{1,3}$, in which the
$\omega_{1,3}$ oscillation has been removed.  In these equations, if
we multiply $\tau_1$ and $\tau_3$ by an overall constant factor $K$,
after some algebra it can be shown that the $A_{1,3}$ equations are
invariant if we rescale $A_1 \to A_1/\sqrt{K}$, $A_3 \to
A_3/\sqrt{K}$, rescale time $t \to K t$, and rescale the input $s_{1+}
\to s_{1+} / K$ [which happens automatically for the critical power by
\eqref{s1crit}].  Note also that the conversion efficiency
$|s_{3-}/s_{1+}|^2 = (2/\tau_3) |A_3 / s_{1+}|^2$ is also invariant
under this rescaling by $K$.  That is, the powers and the timescales of
the dynamics change if you change the lifetimes, unsurprisingly, but
the steady states, stability, etcetera (as investigated in the next
section) are unaltered.

\section{Stability and dynamics}
\label{sec:stability}

\begin{figure}[t]
 \centering
 \includegraphics[width=0.5\textwidth]{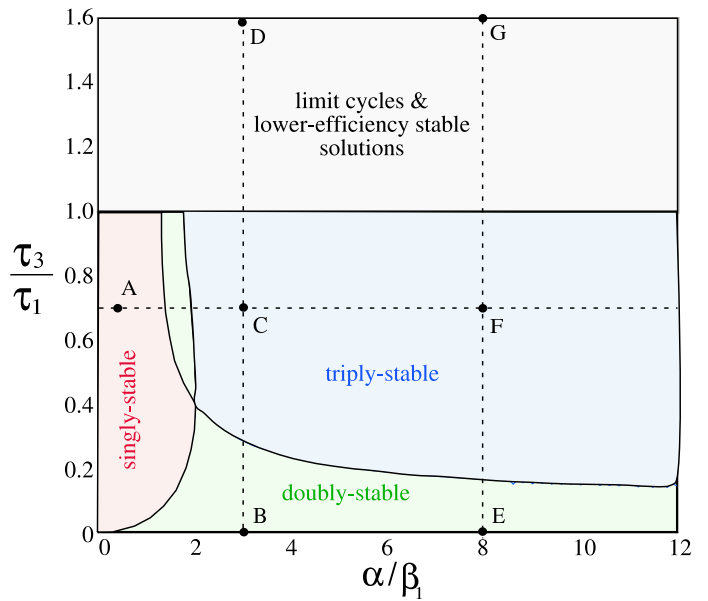}
 \caption{Phase diagram of the nonlinear dynamics of the
 doubly-resonant nonlinear harmonic generation system from
 \figref{schematic} as a function of the relative cavity lifetimes
 ($\tau_3/\tau_1 = 3Q_3/Q_1$) and the relative strength of SPM/XPM
 vs. harmonic generation ($\alpha/\beta_1$) for input power equal to
 the critical power for 100\% efficiency.  For $\tau_3 < \tau_1$ there
 is always one \emph{stable} 100\%-efficiency solution, and for nonzero
 $\alpha$ the system may have additional stable solutions.  For
 $\tau_3 > \tau_1$ the 100\%-efficiency solution becomes unstable, but
 there are limit cycles and lower-efficiency stable solutions.
 Various typical points A--G in each region are labeled for reference
 in the subsequent figures.}
 \label{fig:phasespace}
\end{figure}

Given the steady-state solutions (the roots), their stability is
determined by linearizing the original equations around these points
to a first-order linear equation of the form $d\vec{x}/dt = A\vec{x}$;
a stable solution is one for which the eigenvalues of $A$ have
negative real parts (leading to solutions that decay exponentially
towards the fixed point)~\cite{Tabor89}.

The results of this fixed-point and stability analysis are shown in
\figref{phasespace} as a ``phase diagram'' of the system as a function
of the relative lifetimes $\tau_3 / \tau_1 = 3 Q_3 / Q_1$ and the
relative strength of self-phase-modulation vs. four-wave mixing
$\alpha/\beta_1$.  Our original 100\%-efficiency solution is always
present, but is only stable for $\tau_3 < \tau_1$ and becomes unstable
for $\tau_3> \tau_1$. The transition point, $\tau_3 = \tau_1$,
corresponds to equal energy $|a_1|^2 = |a_3|^2$ in the fundamental and
harmonic mode at the critical input power.  The unstable region
corresponds to $|a_3|^2 > |a_1|^2$ (and the down-conversion term is
stronger than the up-conversion term)---intuitively, this solution is
unstable because, if any perturbation causes the energy in the
harmonic mode to decrease, there is not enough pumping from
up-conversion to bring it back to the 100\%-efficiency
solution. Conversely, in the stable $|a_3|^2 < |a_1|^2$ ($\tau_3 <
\tau_1$) regime, the higher-energy fundamental mode is being directly
pumped by the input and can recover from perturbations.  Furthermore,
as $\alpha/\beta_1$ increases, additional lower-efficiency stable
solutions are introduced, resulting in regimes with two (doubly
stable) and three (triply stable) stable fixed points.  These
different regimes are explored in more detail via bifurcation diagrams
below, and the excitation of the different stable solutions is
considered in the next section.

\begin{figure}[t]
 \centering
 \includegraphics[width=0.5\textwidth]{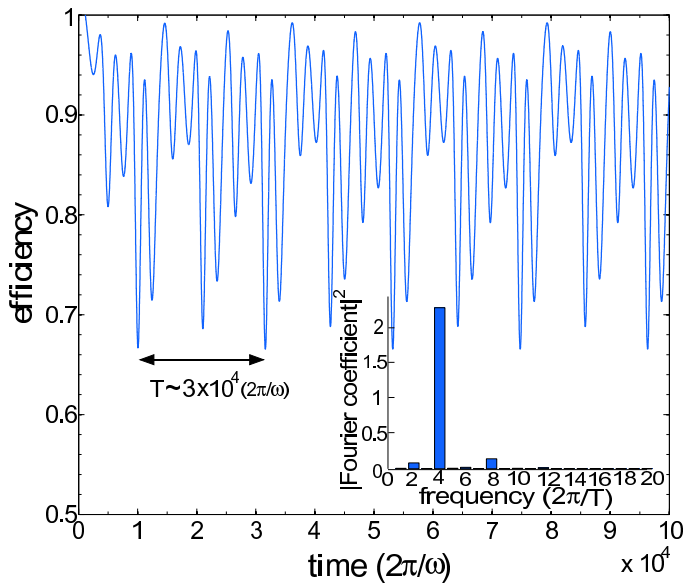}
 \caption{An example of a limit-cycle solution, with a periodically
 oscillating harmonic-generation efficiency as a function of time,
 corresponding to point~D in \figref{phasespace}.  Perturbations in
 the initial conditions produce only phase shifts in the asymptotic
 cycle.  Here, the limit cycle has a period of around $3\times10^4$
 optical cycles.  \emph{Inset:} Square of Fourier amplitudes (arbitary units) for each harmonic component of the limit cycle in the Fourier-series expansion of the $|A_3|$.}
 \label{fig:limcycle}
\end{figure}

For $\tau_3 > \tau_1$, the 100\%-efficiency solution is unstable, but
there are lower-efficiency steady-state solutions and also another
interesting phenomenon: limit cycles.  A limit cycle is a stable
oscillating-efficiency solution, one example of which (corresponding
to point~D in \figref{phasespace}) is plotted as a function time in
\figref{limcycle}.  (In general, the existence of limit cycles is
difficult to establish analytically~\cite{Tabor89}, but the phenomenon
is clear in the numerical solutions as a periodic oscillation
insensitive to the initial conditions).  In fact, as we shall see
below, these limit cycles result from a ``Hopf bifurcation,'' which is
a transition from a stable fixed point to an unstable fixed point and
a limit cycle~\cite{Strogatz94}. In this example at point~D, the efficiency
oscillates between roughly 66\% and nearly 100\%, with a period of
several thousand optical cycles. As a consequence of the time sacling described in the last paragraph of the previous section, the period of such limit cycles is proportional to the $\tau$'s. If the frequency $\omega_1$ were
$1.55\,\mu$m, for a $Q_1$ of $500$ optical cycles, this limit cycle would have a frequency of around $70$~GHz, forming an interesting type of optical ``clock'' or oscillator.  Furthermore, the oscillation is not sinusoidal and
contains several higher harmonics as shown in the inset of
\figref{limcycle}; the dominant frequency component in this case is the fourth harmonic ($\sim 280$~GHz), but different points in the phase diagram yield limit cycles with different balances of Fourier components.

\begin{figure}[t]
 \centering
\includegraphics[width=0.5\textwidth]{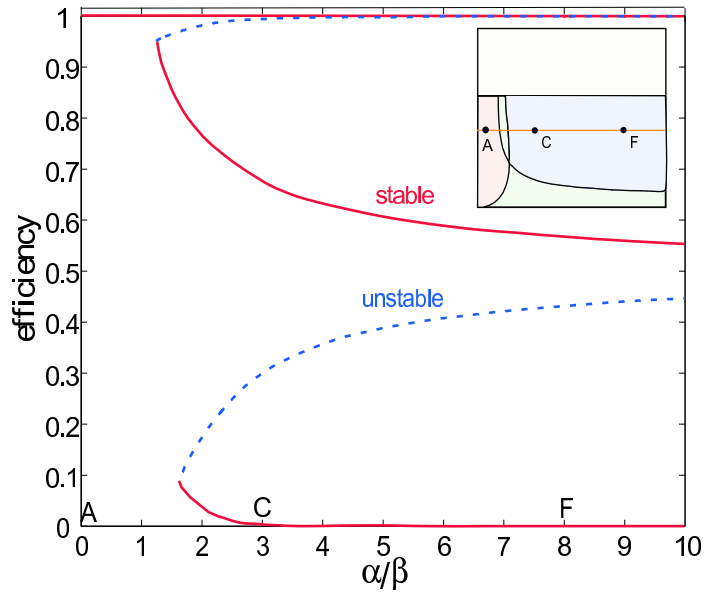}
 \caption{Bifurcation diagram showing the harmonic-generation
 efficiency of the stable (solid red lines) and unstable (dashed blue
 lines) steady-state solutions as a function of $\alpha/\beta_1$ for a
 fixed $\tau_3/\tau_1 = 0.7$, corresponding to the line ACF in
 \figref{phasespace} (see inset). The input power is the critical power
 $P_\mathrm{crit}$, so there is always a 100\%-efficiency stable
 solution, but as $\alpha/\beta_1$ increases new stable and unstable
 solutions appear at lower efficiencies.}
 \label{fig:ACF}
\end{figure}

\begin{figure*}[t]
 \centering
 \includegraphics[width=0.98\textwidth]{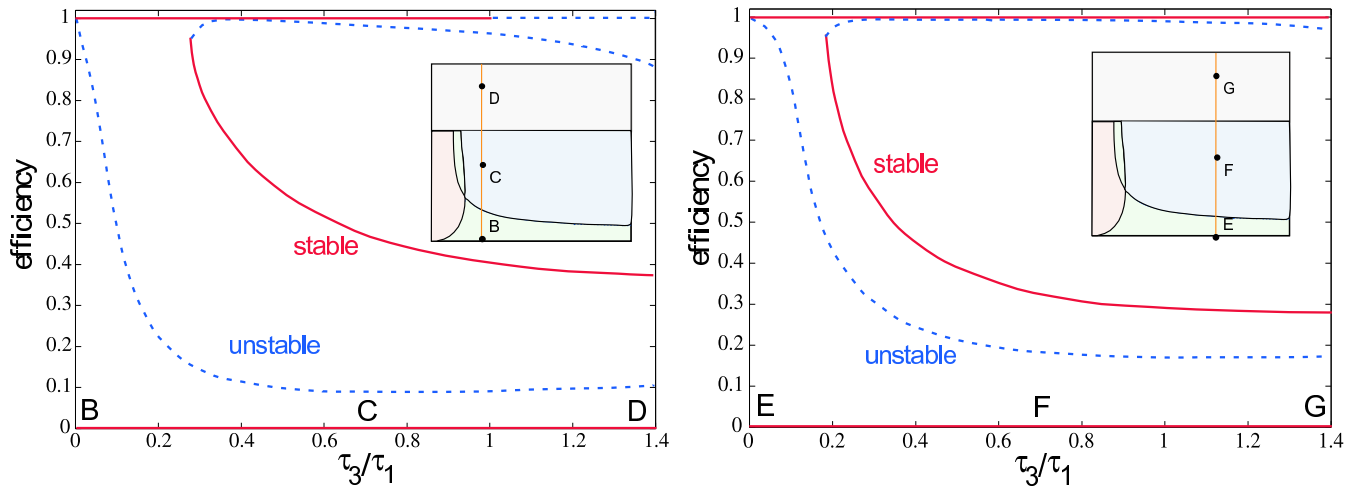}
 \caption{Bifurcation diagram showing the harmonic-generation
 efficiency of the stable (solid red lines) and unstable (dashed blue
 lines) steady-state solutions as a function of $\tau_3/\tau_1$ for a
 fixed $\alpha/\beta_1 = 3$ (left) or $= 8$ (right), corresponding to
 the lines BCD or EFG, respectively, in \figref{phasespace} (see
 insets). The input power is the critical power $P_\mathrm{crit}$, so
 there is always a 100\%-efficiency steady-state solution, but it
 becomes unstable for $\tau_3 > \tau_1$ (a Hopf bifurcation leading to
 limit cycles as in \figref{limcycle}).}
  \label{fig:BCD-ECG}
\end{figure*}

\begin{figure*}[t]
 \centering
 \includegraphics[width=\textwidth]{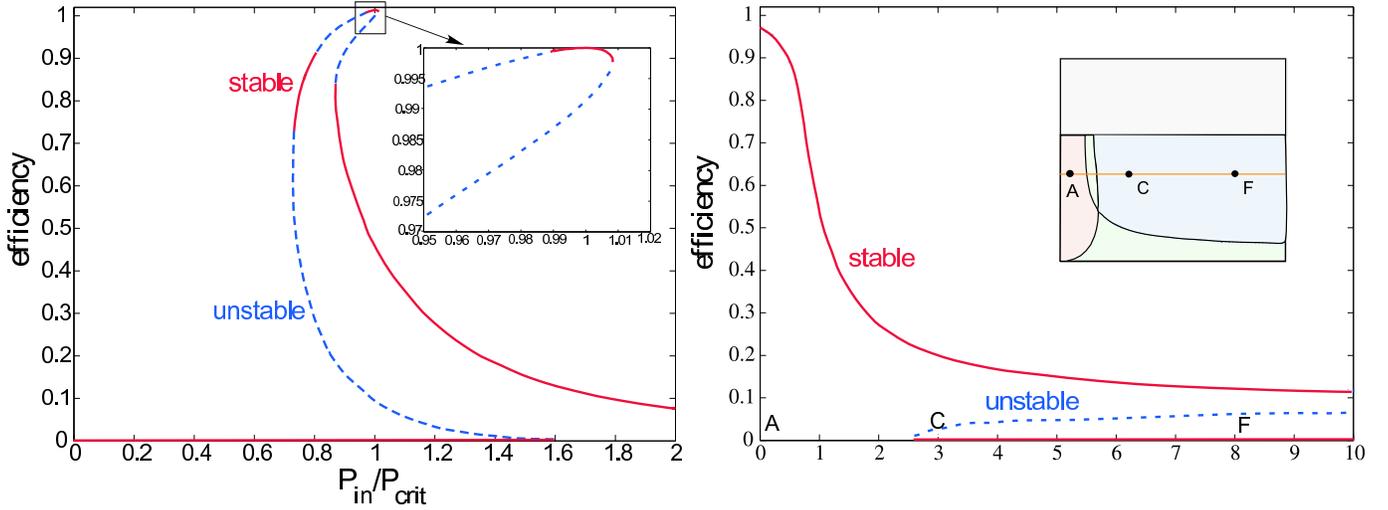}
 \caption{\emph{Left:} Bifurcation diagram showing the
 harmonic-generation efficiency of the stable (solid red lines) and
 unstable (dashed blue lines) steady-state solutions as a function of
 input power $P_\mathrm{in}/P_\mathrm{crit}$ at fixed $\alpha/\beta_1 =
 3$ and $\tau_3/\tau_1 = 0.7$, corresponding to point~C in
 \figref{phasespace}; the inset shows an enlarged view of the
 high-efficiency solutions.  \emph{Right:} Bifurcation diagram as a
 function of $\alpha/\beta_1$ for fixed $P_\mathrm{in}/P_\mathrm{crit}
 = 1.35$ and fixed $\tau_3/\tau_1 = 0.7$; in this case, because it is
 not at the critical power, there are no 100\%-efficiency solutions.}
 \label{fig:P-bifurcation}
\end{figure*}

To better understand the phase diagram of \figref{phasespace}, it is
useful to plot the efficiencies of both the stable and unstable
solutions as a function of various parameters.  Several of these
bifurcation diagrams (in which new fixed points typically appear in
stable/unstable pairs) are shown in \figrefrange{ACF}{P-bifurcation}.
To begin with, \figreftwo{ACF}{BCD-ECG} correspond to lines connecting
the labeled points ACF, BCD, and ECG, respectively, in
\figref{phasespace}, showing how the stability changes as a function
of $\alpha/\beta_1$ and $\tau_3/\tau_1$.  \Figref{ACF} shows how first
one then two new stable fixed points appear as $\alpha/\beta_1$ is
increased, one approaching zero efficiency and the other closer to
50\%.  Along with these two stable solutions appear two unstable
solutions (dashed lines).  (A similar looking plot, albeit inverted,
can be found in \citeasnoun{Felber76} for SPM-coupled closely-spaced
resonances.)  In particular, the fact that one of the unstable
solutions approaches the 100\%-efficiency stable solution causes the
latter to have a smaller and smaller basin of attraction as
$\alpha/\beta_1$ increases, making it harder to excite as described in
the next section.  The next two plots, in \figref{BCD-ECG}, both show
the solutions with respect to changes in $\tau_{3}/\tau_1$ at two
different values of $\alpha/\beta_1$. They demonstrate that at
$\tau_1=\tau_3$, a Hopf bifurcation occurs where the 100\%-efficiency
solution becomes unstable for $\tau_3 \geq \tau_1$ and limit cycles
appear, intuitively seeming to ``bounce between'' the two nearby unstable
fixed points. (The actual phase space is higher dimensional, however, so
the limit cycles are not constrained to lie strictly between the
efficiencies of the two unstable solutions.)  It is worth to note that the
remaining nonzero-efficiency stable solution (which appears at a
nonzero $\tau_3/\tau_1$) becomes less efficient as $\tau_3/\tau_1$
increases.

The above analysis and results were for the steady-state-solutions
when operating at the critical input power to obtain a
100\%-efficiency solution. However, one can, of course, operate with a
different input power---although no other input power will yield a
100\%-efficient steady-state solution, different input powers may
still be useful because (as noted above and in the next section) the
100\%-efficiency solution may be unstable or practically unattainable.
\Figref{P-bifurcation}(left) is the bifurcation diagram with respect
to the input power $P_\mathrm{in}/P_\mathrm{crit}$ at fixed
$\alpha/\beta_1$ and fixed $\tau_3/\tau_1$, corresponding to point~C
in \figref{phasespace}.  This power bifurcation diagram displays a
number of interesting features, with the steady-state solutions
transitioning several times from stable to unstable and vice versa.
As we will see in the next section, the stability transitions in the
uppermost branch are actually supercritical (reversible) Hopf
bifurcations to/from limit cycles.  Near the critical power, there is
only a small region of stability of the near-100\%-efficiency
solution, as shown in the inset of \figref{P-bifurcation}(left). In
contrast, the lower-efficiency stable solutions have much larger
stable regions of the curve while still maintaining efficiencies
greater than 70\% at low powers comparable to $P_\mathrm{crit} \sim
V/Q^2$, which suggests that they may be attractive regimes for
practical operation when $\alpha/\beta_1$ is not small.  This is
further explored in the next section, and also by
\figref{P-bifurcation}(right) which shows the bifurcation diagram
along the line ACF in \figref{phasespace} [similar to \figref{ACF}],
but at 135\% of the critical input power.  For this higher power, the
system becomes at most doubly stable as $\alpha/\beta_1$ is increased,
and the higher-efficiency stable solution becomes surprisingly close
to 100\% as $\alpha/\beta_1 \to 0$.

\section{Exciting high-efficiency solutions}
\label{sec:exciting}

\begin{figure}[t]
 \centering
 \includegraphics[width=0.48 \textwidth]{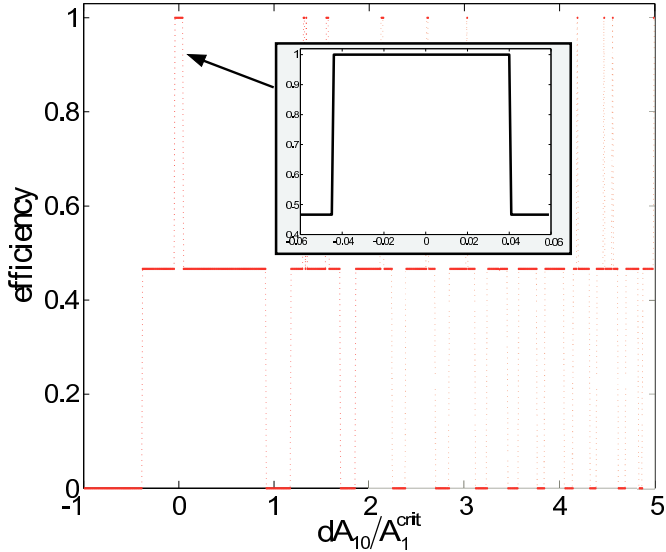}
 \caption{Asymptotic steady-state efficiency at point~C
 (triply-stable) in the phase diagram (\figref{phasespace}), with the
 initial conditions perturbed from the 100\%-efficiency stable
 solution.  The initial amplitudes $A_{10}$ and $A_{30}$ are perturbed
 by $\delta A_{10}$ and $\delta A_{30}$, respectively, with $\delta
 A_{10} / A_{1}^\mathrm{crit} = \delta A_{30} /
 A_{3}^{\mathrm{crit}}$.  The oscillation of the steady-state
 efficiency with the perturbation strength is an indication of the
 complexity of the phase space and the shapes of the basins of
 attraction of each fixed point.}
 \label{fig:prim-adet}
\end{figure}

One remaining concern in any multistable system is how to excite the
desired solution---depending on the initial conditions, the system may
fall into different stable solutions, and simply turning on the source
at the critical input power may result in an undesired low-efficiency
solution.  If $\alpha/\beta$ is small enough, of course, then from
\figref{phasespace} the high-efficiency solution is the only stable
solution and the system must inevitably end up in this state no matter
how the critical power is turned on.  Many interesting physical
systems will correspond to larger values of $\alpha/\beta$,
however~\cite{Rodriguez07:OE}, and in this case the excitation problem
is complicated by the existence of other stable solutions.  Moreover,
the basins of attraction of each stable solution may be very
complicated in the phase space, as illustrated by \figref{prim-adet},
where varying the initial cavity amplitudes $A_{1,3}$ from the
100\%-efficiency solution causes the steady state to oscillate in a
complicated way between the three stable solutions (at point~C in
\figref{phasespace}).  We have investigated several solutions to this
excitation problem, and found an ``adiabatic'' excitation technique
that reliably produces the high-efficiency solution without
unreasonable sensitivity to the precise excitation conditions.

%

 \begin{figure}[t]
 \centering
 \includegraphics[width=0.48 \textwidth]{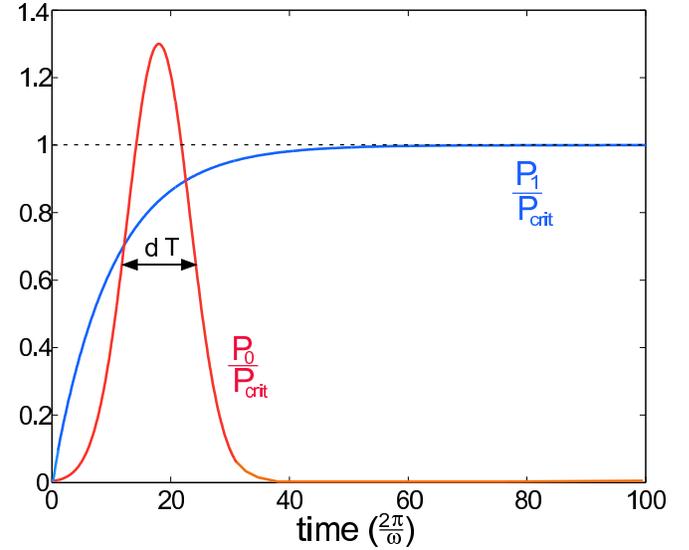}
 \caption{One way of exciting the system into a controlled stable
 solution: the input power is the sum of an exponential turn-on (the
 blue curve, $P_1$) and a Gaussian pulse with amplitude $P_0$ and
 width $\delta T$.  The amplitude $P_0$ is altered to control which
 stable solution the system ends up in.}
 \label{fig:prim-pulse}
 \end{figure}

 \begin{figure*}[t]
 \centering
 \includegraphics[width=0.98\textwidth]{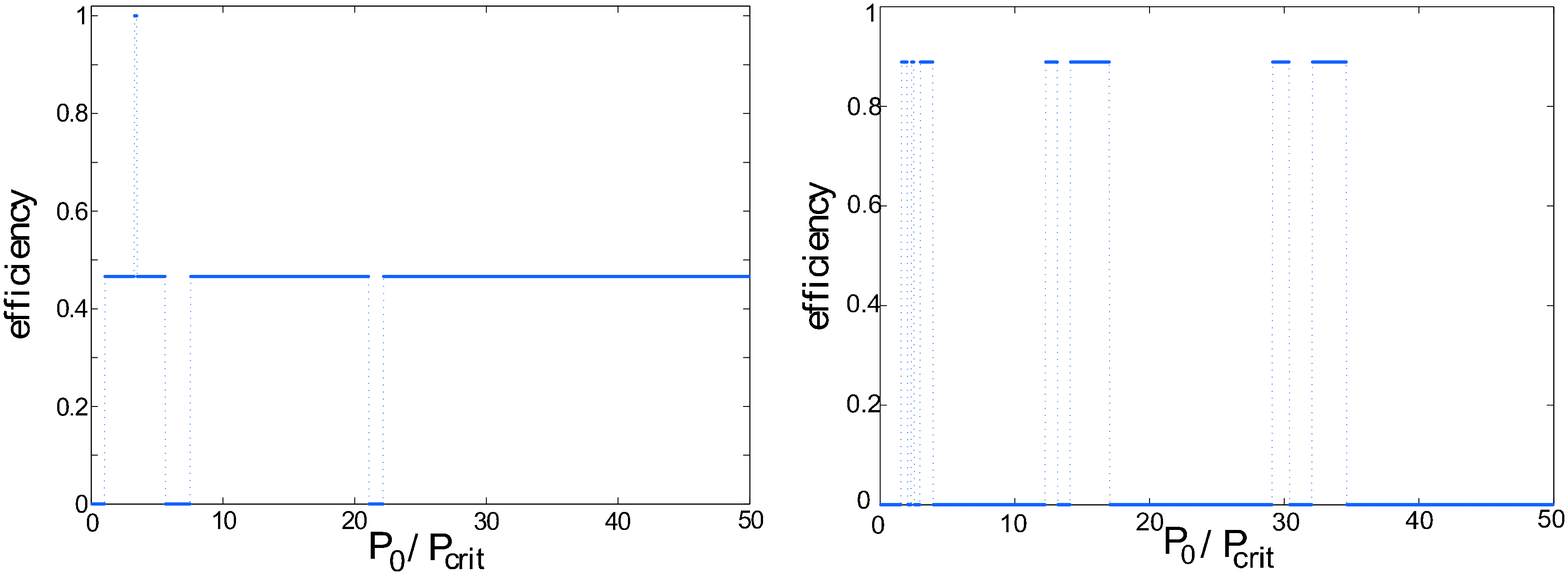}
 \caption{\emph{Left:} Steady-state efficiency at point~C in
 \figref{phasespace} as a function of the transient input-pulse power
 $P_0$ from \figref{prim-pulse}, showing how all three stable
 solutions can be excited by an appropriate input-pulse amplitude.
 \emph{Right:} Same, but for an asymptotic input power $P_1 \approx 0.8
 P_\mathrm{crit}$, for which the maximum efficiency is $\approx 90\%$
 from \figref{P-bifurcation}(right), but is easier to excite.}
 \label{fig:prim}
 \end{figure*}

First, we considered a simple technique similar to the one described
in \citeasnoun{Soljacic02:bistable} for exciting different solutions
of a bistable filter: as shown in \figref{prim-pulse}, we ``turn on''
the input power by superimposing a gradual exponential turn-on
(asymptoting to $P_1 = P_\mathrm{crit}$) with a Gaussian pulse of
amplitude $P_0$ and width $\delta T$.  The function of the initial
pulse is to ``kick'' the system into the desired stable solution.  We
computed the eventual steady-state efficiency (after all transient
effects have disappeared) as a function of the pulse amplitude $P_0$
at point~C in \figref{phasespace}, where there are three stable
solutions.  The results are shown in \figref{prim}(left), and indeed
we see that all three stable solutions from point~C in
\figref{ACF}: one at near-zero efficiency, one at around 47\%
efficiency, and one at 100\% efficiency.  Unfortunately, the 100\%
efficiency solution is obviously rather difficult to excite, since it
occurs for only a very narrow range of $P_0$ values.  One approach to
dealing with this challenge is to relax the requirement of 100\%
efficiency (which will never be obtained in practice anyway due to
losses), and operate at a power $P_1 < P_\mathrm{crit}$.  In
particular, \figref{P-bifurcation}(left) shows that there is a much
larger stable region for $P_1 \approx 0.8 P_\mathrm{crit}$ with
efficiency around 90\%, leading one to suspect that this solution may
be easier to excite than the 100\%-efficiency solution at $P_1 =
P_\mathrm{crit}$.  This is indeed the case, as is shown in
\figref{prim}(right), plotting efficiency vs. $P_0$ at point~C with
$P_1 \approx 0.8 P_\mathrm{crit}$.  In this case, there are only two stable
solutions, consistent with \figref{P-bifurcation}(left), and there are
much wider ranges of $P_0$ that attain the high-efficiency ($\approx
90\%$) solution.

There are also many other ways to excite the high-efficiency solution
(or whatever steady-state solution is desired).  For example, because
the cavity is initially detuned from the input frequency, as described
in \secref{theory}, much of the initial pulse power is actually
reflected during the transient period, and a more efficient solution
would vary the pulse frequency in time to match the cavity frequency
as it detunes.  One can also, of course, vary the initial pulse width
or shape, and by optimizing the pulse shape one may obtain a more robust solution.

\begin{figure*}[t]
 \centering
 \includegraphics[width=0.98\textwidth]{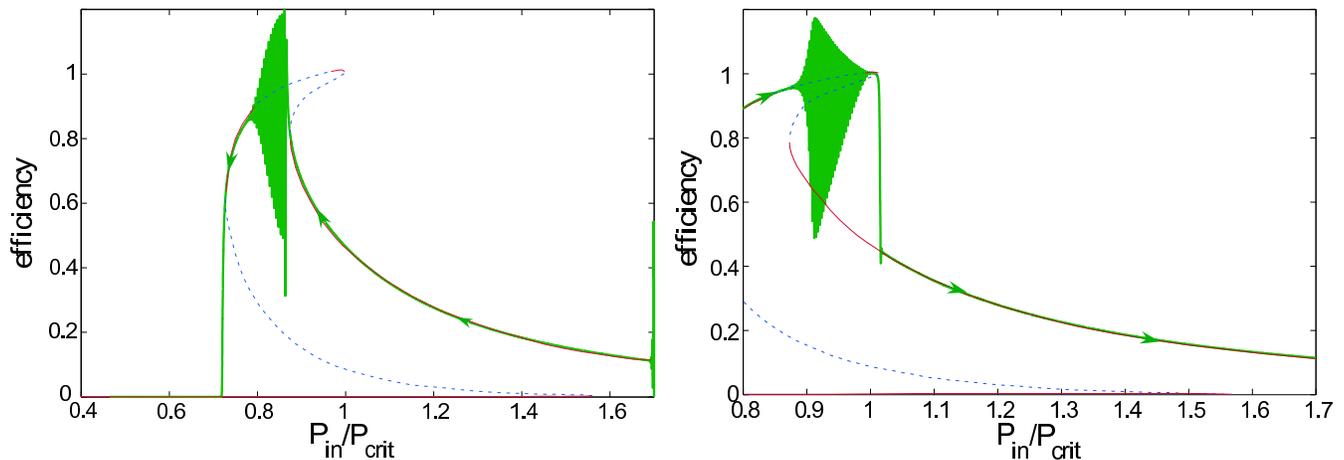}
 \caption{\emph{Left:} Black line with arrows indicates instantaneous
 ``efficiency'' (harmonic output power / input power) as the input
 power is slowly decreased, starting at a power $\approx 1.7
 P_\mathrm{crit}$.  For comparison, \figref{P-bifurcation}(left) is
 superimposed as solid-red and dashed-blue lines.  The solution
 ``adiabatically'' follows a steady state until the steady state
 becomes unstable, at which point it enters limit cycles, and then
 returns to a high-efficiency steady state, and finally goes drops to
 a low-efficiency steady-state if the power is further decreased.
 \emph{Right:} Similar, but here the power is \emph{increased}
 starting at the high-efficiency steady state solution for $P <
 P_\mathrm{crit}$.  In this case, it again enters limit cycles, but
 then it returns to a high-efficiency steady-state solution as the
 power is further increased, eventually reaching the 100\%-efficiency
 stable solution.  If the power is further increased, it drops
 discontinuously to the remaining lower-efficiency steady-state stable
 solution.}
 \label{fig:exciting}
 \end{figure*}

In particular, one can devise a different (constant-frequency) input
pulse shape that robustly excites the high-efficiency solution,
insensitive to small changes in the initial conditions, by examining
the power-bifurcation diagram in \figref{P-bifurcation}(left) in more
detail.  First, we observe that input powers $\gtrsim 1.45
P_\mathrm{crit}$ have only one stable solution, meaning that this
stable solution is excited regardless of the initial conditions or the
manner in which the input power is turned on.  Then, if we slowly
decrease the input power, the solution must ``adiabatically'' follow
this stable solution in the bifurcation diagram until a power $\approx
0.95 P_\mathrm{crit}$ is reached, at which point that stable solution
disappears.  In fact, by inspection of \figref{P-bifurcation}(left),
at that point there are \emph{no} stable solutions, and solution jumps
into a limit cycle. If the power is further decreased, a
high-efficiency stable solution reappears and the system must drop
into this steady state (being the only stable solution at that point).
This process of gradually decreasing the power is depicted in
\figref{exciting}(left), where the instantaneous ``efficiency'' is
plotted as a function of input power, as the input power is slowly
decreased.  (The efficiency can exceed unity,because we are plotting
instantaneous output vs.~input power, and in the limit-cycle
self-pulsing solution the output power is concentrated into pulses
whose peak can naturally exceed the average input or output power.)
Already, this is an attractive way to excite a high-efficiency ($>
90\%$) solution, because it is insensitive to the precise manner in
which we change the power as long as it is changed slowly
enough---this rate is determined by the lifetime of the cavity, and
since this lifetime is likely to be sub-nanosecond in practice, it is
not difficult to change the power ``slowly'' on that timescale.
However, we can do even better, once we attain this high-efficiency
steady state, by then \emph{increasing} the power adiabatically. As we
increase the power, starting from the high-efficiency steady-state
solution below the critical power, the system first enters limit-cycle
solutions when the power becomes large enough that the stable solution
disappears in \figref{P-bifurcation}(left).  As we increase the power
further, however, we observe that these limit cycles \emph{always}
converge adiabatically into the 100\%-efficiency solution when $P\to
P_\mathrm{crit}$.  This process is shown in \figref{exciting}(right).
What is happening is actually a supercritical Hopf bifurcation at the
two points where the upper branch changes between stable and unstable:
this is a reversible transition between a stable solution and a limit
cycle (initially small oscillations, growing larger and larger away
from the transition).  This is precisely what we observe in
\figref{exciting}, in which the limit cycle amplitudes become smaller
and smaller as the stable solutions on either side of the upper branch
are approached, leading to the observed reversible transitions between the two.
The important fact is that, in this way, by first decreasing and then
increasing the power to $P_\mathrm{crit}$, one always obtains the
100\%-efficiency solution regardless of the precise details of how the
power is varied (as long as it is ``slow'' on the timescale of the
cavity lifetime).

\section{Concluding Remarks}

We have shown that a doubly-resonant cavity not only has
high-efficiency harmonic conversion solutions for low input power, as
in our previous work~\cite{Rodriguez07:OE}, but also exhibits a number
of other interesting phenomena.  We showed under what conditions the
high-efficiency solution is stable, how to compensate for self-phase
modulation, the existence of different regimes of multistable
solutions and limit cycles controlled by the parameters of the system
and by the input power, and how to excite the desired high-efficiency
solution.  Although we did not observe chaos, it seems possible that
this may be obtained in future work for other parameter regimes,
e.g. for pulsed input power as was observed in the $\chitwo$
case~\cite{Drummond80}.  These dynamical phenomena depend only on
certain dimensionless quantities $\alpha/\beta$, $\omega_3/\omega_1$,
$\tau_3/\tau_1$, and $s_{1+} / s_{1+}^\mathrm{crit}$, although the
overall power and time scales depend upon the dimensionful quantities
$\tau_{1,3}$ and so on.

All of the calculations in this paper were for an idealized lossless
system, as our main intention was to examine the fundamental dynamics
of these systems rather than a specific experimental realization.
However, we have performed preliminary calculations including both
linear losses (such as radiation or material absorption) and nonlinear
two-photon absorption, and we find that these losses do not
qualitatively change the observed phenomena.  One still obtains
multistability, limit cycles, bifurcations, and so on, merely at
reduced peak efficiency depending on the strength of the losses.  In a
future manuscript, we plan to explore these effects in more detail in
realistic material settings, and propose specific geometries to obtain
the requisite doubly-resonant cavities.  In particular, to obtain
widely-spaced resonant modes $\omega$ and $3\omega$ in a nanophotonic
(wavelength-scale) context (as opposed to macroscopic Fabry-Perot
cavities with mirrors), the most promising route seems to be a ring
resonator of some sort~\cite{Saleh91}, rather than a photonic
crystal~\cite{JoannopoulosJo08-book} (since photonic band gaps at
widely separated frequencies are difficult to obtain in two or three
dimensions).  Although such a cavity will naturally support more than
the two $\omega_1$ and $\omega_3$ modes, only two of the modes will be
properly tuned to achieve the resonance condition for strong nonlinear
coupling.

Finally, we should mention that similar phenomena should also arise in
doubly and triply resonant cavities coupled nonlinearly by
sum/difference frequency generation (for $\chitwo$) or four-wave
mixing (for $\chithree$).  The advantage of this is that the coupled
frequencies can lie closer together, imposing less stringent materials
constraints and allowing the cavity to be confined by narrow-bandwidth
mechanisms such as photonic band gaps~\cite{JoannopoulosJo08-book}, at
the cost of a more complicated cavity design.

\section*{Acknowledgements}

This research was supported in part by the Army Research Office through the Institute for Soldier Nanotechnologies under Contract No. W911NF-07-D-0004. This work was also supported in part by a Department of Energy (DOE)
Computational Science Fellowship under grant DE--FG02-97ER25308 (AWR).
We are also grateful to A. P. McCauley at MIT and S. Fan at Stanford for
helpful discussions.


\end{document}